\documentclass[aps,preprint]{revtex4}%
\usepackage{amsfonts}
\usepackage{amsmath}
\usepackage{amssymb}
\usepackage{graphicx}%
\providecommand{\U}[1]{\protect\rule{.1in}{.1in}}
\begin{document}
\preprint{ }
\title[Gauge nonlocality]{Gauge nonlocality in planar quantum-coherent systems}
\author{K. Moulopoulos}
\email{cos@ucy.ac.cy}
\affiliation{University of Cyprus, Department of Physics, 1678 Nicosia, Cyprus}

\begin{abstract}
It is shown that a system with quantum coherence can be nontrivially affected
by adjacent magnetic or adjacent time-varying electric field regions, with
this proximity (or remote) influence having a gauge origin. This is implicit
(although overlooked) in numerous works on extended systems with inhomogeneous
magnetic fields (with either conventional or Dirac materials) but is generally
plagued with an apparent gauge ambiguity. The origin of this annoying feature
is explained and it is shown how it can be theoretically removed, leading to
macroscopic quantizations (quantized Dirac monopoles, integral quantum Hall
effect, quantized magnetoelectric phenomena in topological insulators). Apart
however from serving as a theoretical probe of macroscopic quantizations,
there are cases (experimental conditions, clarified here) when this
\textquotedblleft gauge nonlocality\textquotedblright\ does not really suffer
from any ambiguity: an apparently innocent gauge transformation corresponds to
real change in physics of a companion system in higher dimensionality, that
leads to physical momentum transfers to our own system. This nonlocality,
together with the associated \textquotedblleft proximity\textquotedblright\ or
remote effects are then real and lead to the remarkable possibility of
inducing topological phenomena from outside our system (which always remains
field-free and can even reside in simply-connected space). Specific procedures
are then proposed to experimentally detect such types of nonlocal effects and
exploit them for novel applications. General consequences in solid state
physics (such as the first violation of Bloch theorem in a field-free quantum
periodic system) are pointed out, and formal analogies with certain high
energy physics phenomena (axions, $\theta$-vacua and some types of Gribov
ambiguities), as well as with certain largely unexplored phenomena in
mechanics and in thermodynamics, are noted.

\end{abstract}
\volumeyear{2013}
\volumenumber{number}
\issuenumber{number}
\eid{identifier}
\date[October 18, 2013]{}
\startpage{1}
\endpage{ }
\maketitle

\section{\bigskip Introduction}

A universal dynamical nonlocality in two-dimensional (2D) quantum systems is
reported that is of a gauge nature.\textbf{ }This is done by demonstrating,
using a gauge argument, that interesting and nontrivial physics may occur
inside a 2D system with quantum coherence that is \textbf{nearby} magnetic or
time-dependent electric field regions\textbf{. }First, a remote (or proximity)
dynamic influence of such fields on adjacent regions in flat 2D space is shown
to be a natural consequence of \textbf{hidden} Aharonov-Bohm (AB) effects
(magnetic or electric)\cite{AB} combined with the absence of magnetic
monopoles\cite{Dirac} in higher dimensionality (3D). This proximity effect is
here rigorously shown to exist and to affect numerous results in the
literature on extended and open arrangements with \textbf{inhomogeneous}
magnetic fields (involving either conventional or Dirac materials), \textit{if
quantum coherence parallel to the interfaces is taken into account}. This has
apparently not been noted in previous works -- the reason possibly being that
there generally remains a gauge ambiguity. It is shown that this ambiguity in
the plane is actually due to the richer physics of a companion system in 3D
that reduces to our 2D system in an appropriate limit. Under such a limiting
procedure (and under certain experimental conditions) it is shown that, in
fact, this \textquotedblleft gauge nonlocality\textquotedblright\ does
\textit{not} really suffer from any ambiguity$:$ an apparently innocent gauge
transformation actually corresponds to real change in physics, due to
nonequivalent displacements of the 3D companion relative to our 2D system (but
all of them producing the same result in the proper planar limit) $-$ and this
involves physical momentum transfers to our remote system, with all the
physical consequences of a genuine nonlocal effect. This nonlocality then has
important applications to extended systems with adjacent time-dependent
electric fields, or with adjacent adiabatically varying magnetic fields (in
their intensity or in their placement in 3D space); these lead to the
possibility of manufacture of interesting quantum devices that exploit the
above proximity influence (i.e. of spacetime electric fluxes) to induce
topological phenomena from \textbf{outside} the system $-$ the simplest
example being an electric flux-driven charge pumping in a modification of the
well-known Laughlin's gauge argument\cite{Laughlin} that is usually invoked
for the explanation of the Integral Quantum Hall Effect (IQHE). From analysis
of the 3D companion system it is made clear that the above proximity effects
are not only real (can be realized experimentally), but they also give the
possibility \textbf{(A)} of an easier experimental detection of AB effects (in
a simply-connected system and without enclosed fluxes), and \textbf{(B)} of
having the first example of a planar \textit{field-free} periodic system
(crystal) that actually violates Bloch's theorem (due to the hidden AB effect
caused by the 3D companion). We propose specific ways through which an
experimentalist can measure effects related to the above, hinting at expected
behaviors in a conventional 2D solid state system (i.e. with parabolic energy
spectrum), but also in graphene and topological insulator surfaces (examples
of materials with low-energy linear energy spectrum). However, in a strict
planar world, with complete lack of information on the 3D companion, this
ambiguity may show up (actually reflecting our ignorance). It can then be
theoretically removed when certain adjacent fluxes are properly quantized,
which immediately suggests a natural way to eliminate the artificial effect
for confined systems (closed manifolds), and we propose this (enforcement of
elimination of the ambiguity, through quantization of nearby fluxes) as a
criterion of proper behavior. We show that this has direct applicability even
to cases when (effective) magnetic monopoles are present; the same criterion
then directly leads to the quantization of certain macroscopic quantities, and
this in turn leads to topological quantization of charge and response
functions in a wide range of systems of current interest without further gauge
considerations. Examples include the standard Dirac quantization of magnetic
monopoles\cite{Dirac}, and $-$ by additionally invoking axion
electrodynamics\cite{Wilczek} $-$ the integral quantization of Hall
conductance in conventional 2D Quantum Hall systems, and also the
\textquotedblleft half-quantization\textquotedblright\ of the recently
discussed quantized magnetoelectric phenomena in surface-gapped 3D
time-reversal-symmetric topological insulators (by also demonstrating that
this half-quantization basically reflects the Witten effect\cite{Witten}).
Finally, connections are noted with certain high-energy physics phenomena that
seem to have a formal similarity ($\theta$-vacua, and some types of Gribov
ambiguities), as well as with certain areas in mechanics and in thermodynamics
that are still largely unexplored; a mapping is also briefly mentioned to
general spin-related phenomena, through boosts to properly moving frames,
giving the possibility of studying nontrivial spin-physics by starting from
purely orbital considerations, although a serious look at spin-related
phenomena (including spin-orbit interactions) in this new framework is
reserved for a future note.

\section{The system}

Consider a flat rectangle (strip) of horizontal length $L$ in the ($xy$)-plane
with periodic boundary conditions (pbc) along $L$ (in the $x$-direction),
consisting of two adjacent (up and down in the $y$-direction) parts, again
strips of length $L$, the one on top being empty of fields or scalar
potentials (the \textquotedblleft white\textquotedblright\ area) and the one
at the bottom penetrated by a perpendicular magnetic field $B$ (the
\textquotedblleft dark\textquotedblright\ area). We start with a static and
uniform $B$ (although this will be relaxed later), and we first consider a
nonrelativistic quantum particle (of mass $m$ and charge $e$) that moves
\textbf{only} inside the upper white area (i.e. the two areas are separated by
an appropriate scalar potential wall, so that the lower dark (magnetic) area
is totally inaccessible to the particle). Let us then set the origin $y=0$ at
the floor of the dark area, i.e. take $(0,0)$ at the bottom left corner of the
dark (magnetic) strip, the separating wall being at $y=d_{1}$, and the ceiling
of the white area being at $y=d_{2}$ (which, for simplicity, we also consider
to be impenetrable). The particle is therefore confined in the $y$-direction
by the walls at $y=d_{1}$ and $y=d_{2}$, with periodic boundary conditions in
the $x$-direction, and feels no magnetic field $-$ the field $B$ being only in
the adjacent dark \textquotedblleft forbidden\textquotedblright\ area, that
lies below the particle's white strip. The usual procedure to solve this
rather trivial problem, especially for the $B=0$ case, would be to work in the
gauge $\mathbf{A}=0$ everywhere inside the white region: eigenfunctions are
then of the form $\Psi(x,y)\thicksim e^{ik_{x}x}\sin k_{y}(y-d_{1})$ (with
$k_{y}=\frac{n_{y}\pi}{d},$ $\ n_{y}=1,2,...$ and $d=d_{2}-d_{1}$, and with
$k_{x}=\sqrt{\frac{2m(\epsilon-\frac{\hbar^{2}k_{y}^{2}}{2m})}{\hbar^{2}}}$
being quantized as $k_{x}=\frac{2\pi}{L}n_{x}$ \ ($n_{x}=0,\pm1,\pm2,...$)),
with the associated energies being therefore \ $\epsilon_{n_{x},n_{y}}%
=\frac{\hbar^{2}}{2m}\frac{\pi^{2}n_{y}^{2}}{d^{2}}+\frac{\hbar^{2}}%
{2m}\left(  \frac{2\pi}{L}\right)  ^{2}n_{x}^{2}$ . Let us now include the
magnetic field $B$ (that is always inside the dark area only) by using a
generalization of the Landau gauge (and a special case of one used by Bawin \&
Burnel long ago\cite{BawinBurnel}), with the origin being as noted above,
namely $\mathbf{A=}-yB\hat{e}_{x}$ \ for $0\leq y\leq d_{1},$ \ and
\ $\mathbf{A=}-d_{1}B\hat{e}_{x}\equiv A_{_{0}}\hat{e}_{x}$ \ for $d_{1}\leq
y\leq d_{2}$; this gauge choice indeed satisfies that $\frac{\partial A_{y}%
}{\partial x}-\frac{\partial A_{x}}{\partial y}$ is $B$ inside and zero
outside the dark region, and $\mathbf{A}$ is continuous at the separating wall
(at $y=d_{1}$). Note that the particle in the white area now feels a nonzero
(although uniform) vector potential, that makes wavefunctions formally pick up
an extra phase factor $e^{i\frac{e}{\hbar c}\int^{x}\mathbf{A\cdot}%
d\mathbf{x}^{\prime}}$ (through a gauge transformation mapping trick, starting
from $\mathbf{A}=0$), so that we now have $\Psi^{(\mathbf{A)}}(x,y)\thicksim
e^{ik_{x}x}e^{i\frac{e}{\hbar c}A_{_{0}}x}\sin\frac{n_{y}\pi}{d}y$ . By then
imposing the pbc in the $x$-direction, we obtain \ $e^{i(k_{x}+\frac{e}{\hbar
c}A_{_{0}}\mathbf{)}L}=1.$ From this, we can determine the new quantized
values of $k_{x}$ and then the energy spectrum, which finally turns out to be \ %

\begin{equation}
\ \epsilon_{n_{x},n_{y}}=\frac{\hbar^{2}}{2m}\frac{\pi^{2}n_{y}^{2}}{d^{2}%
}+\frac{\hbar^{2}}{2m}\left(  \frac{2\pi}{L}\right)  ^{2}(n_{x}-f)^{2},
\end{equation}
with \ $f=\frac{\Phi_{dark}}{\Phi_{0}}$, \ where \ $\Phi_{dark}=Bd_{1}L$ \ is
the total flux through the dark area and $\Phi_{0}=\frac{hc}{e}$ is the flux
quantum. These allowed energies are actually periodic with respect to
$\Phi_{dark}$ (with period $\Phi_{0}$) $-$ as can be seen if, for a given
$\Phi_{dark}$, proper shifting of the integers $n_{x}$ is made $-$ and
whenever $\Phi_{dark}$ happens to be an integral multiple of $\Phi_{0}$, the
global spectrum is equivalent to that corresponding to the absence of the
adjacent $B$ (i.e. to $\Phi_{dark}\mathbf{=}0$, reducing to the one with
$\mathbf{A}=0$ derived earlier).

The key observation is that, although the particle will never enter the dark
area, its energy spectrum, and from this other measurable quantities (i.e.
global electric current $J=-c\frac{\partial\epsilon}{\partial\Phi}$) are seen
to be affected by the adjacent (forbidden) magnetic field $-$ a type of
\textquotedblleft proximity field influence\textquotedblright, and not the
usual AB effect, since the magnetic flux is \textbf{not enclosed} by the
region where the particle resides, but is only adjacent to it. If the origin
of our coordinate system were chosen anywhere below the dark floor, the above
result would seem to be origin-independent. If however we chose the origin to
be, e.g., at the wall separating the two areas, then this \textquotedblleft
effect\textquotedblright\ would go away. (And note that in flat space, change
of origin is equivalent to a gauge transformation $-$ see further on this
later below). We observe therefore a gauge ambiguity. Hence one may well say
that it cannot be a real physical effect; the theory however \textit{does
predict} such an \textquotedblleft artificial\textquotedblright\ effect as a
direct consequence. What is the reason behind it or what is its deeper origin?
And, most importantly, is it ever possible to make any use of it
experimentally? We shall answer in the affirmative, under certain conditions.

An immediate first understanding comes from the appearance of nonlocal terms
in a gauge function\cite{KM1,KM2}, see Appendix B. However, the deeper origin
of this proximity field influence\ in flat space can be revealed through 3D
folding (compatible with the pbc in the $x$-direction): we show in what
follows that the above \textquotedblleft effect\textquotedblright\ actually
originates from the absence of magnetic monopoles anywhere in the embedding 3D
space. Indeed, by folding in the $x$-direction to form a cylinder (by gluing
the opposite vertical sides), the above gauge, now written in cylindrical
coordinates, has only azimuthal component, and it is $\mathbf{A=(}%
0,-zB\mathbf{,}0\mathbf{)}$ in the dark and $\mathbf{A=(}0,-d_{1}%
B\mathbf{,}0\mathbf{)}$ in the white area, with $B$ always denoting the
magnitude of the locally perpendicular field, that has now become the
\textit{radial} component of a total magnetic field $\mathbf{B}$ in 3D space.
It is crucial then to note that this gauge choice leads in 3D to the
additional appearance of a nonzero $B_{z}$ (component of $\mathbf{B}$ parallel
to the cylinder's $z$-axis) that is inhomogeneous (generally $\varrho$- and
$z$-dependent). Indeed, straightforward calculation of the total field
$\mathbf{B}$ produced by the above form of $\mathbf{A}$ (see Appendix A) leads
to $B_{z}=-\frac{zB}{\varrho}$ in the dark area, and $B_{z}=-\frac{d_{1}%
B}{\varrho}$ in the white area. This inhomogeneous $B_{z}$ in all space is
equal to exactly what is needed to give a flux (of this $B_{z}$) through the
ceiling (say at height $z_{2}$) and the floor (say at height $z_{1}$) of any
cylinder (of height $z_{2}-z_{1}$) that overall \textit{cancels out} the flux
(of $B_{\varrho}$) that goes through its curved cylindrical side-surface; and
the flux of $B_{z}$ through the ceiling is also identical to the value of a
horizontal closed integral of the corresponding $\mathbf{A}$ at height $z_{2}$
$-$ this way directly demonstrating that the above cancellation is actually
due to the standard AB effect (since the $B_{z}$-flux \textit{is enclosed} by
the particle's region). What we see here is simply that the total flux passing
through the entire closed cylindrical surface is indeed zero (as demanded by
the volume integral of $\mathbf{\nabla\cdot B}=0$ inside the whole cylinder).
Hence, in the case of $z_{2}=z$ being in the white region, and $z_{1}=0$, the
\textquotedblleft proximity field influence\textquotedblright\ at height $z$
inside the white area is essentially the usual AB effect, since, as noted, a
closed path in the white area encloses the flux of $B_{z}$, which is equal (up
to a sign) to the \textquotedblleft dark flux\textquotedblright, i.e. the flux
of $B$ through the entire dark strip (the contribution from the floor at
$z_{1}=0$ being vanishing). This way \textbf{the dark strip affects indirectly
(}through the companion system in 3D\textbf{) the adjacent white region}.

Seeking an experimental realization, let us momentarily turn to a slightly
different gauge, namely $\mathbf{A=(}0,-\frac{zBR}{\rho}\mathbf{,}0\mathbf{)}$
in the dark area (a gauge discussed earlier\cite{Chryssomalakos}, together
with an actual realistic current distribution $\mathbf{J}$ that produces it)
and $\mathbf{A=(}0,-\frac{d_{1}BR}{\rho}\mathbf{,}0\mathbf{)}$ in the white
area (all this being compatible with our own gauge for $\varrho\rightarrow R$
the radius of the cylinder). This gauge, produced by a $\mathbf{J}%
\propto-BRz\hat{e}_{\varphi}$, can be shown to lead to similar cancellations
and a similar conclusion of influence of remote fields, see Appendix A. But,
more importantly, in both gauges, the value of $B_{z}$ changes with the
location of origin; i.e. in our first choice of gauge, it generally becomes
$B_{z}=-\frac{(z-z_{0})B}{\varrho}$ for $z$ in the dark area, and
$B_{z}=-\frac{(d_{1}-z_{0})B}{\varrho}$ for $z$ in the white area, and this
can be seen as the actual source of the gauge ambiguity noted earlier; it has
to do with a different flux balance (in the overall cancellation) between the
ceiling, the floor and the side-surface of the considered cylinder, and this
will generally give an origin-dependent flux in the ceiling $-$ hence leading
to a $z_{0}$-dependent AB influence at height $z$, and therefore a $y_{0}%
$-dependent proximity influence in the initial flat system. Note that, in both
1st and 2nd choice of gauge, the point $z_{0}$ is always the point (height)
where $B_{z}$ (or $\mathbf{J}$) vanishes (see also ref.\cite{Westerberg},
fig.3, for a related (but simpler) system) $-$ these observations being
important for our later discussion on a relevant experimental setup.

In spite of the above peculiarity however (namely, the extra appearance of a
$B_{z}$ that actually has a vanishing point with a totally arbitrary
location), the crucial property to note is that, when the plane is flat (i.e.
in the limit $R\rightarrow\infty$),\textit{ }\textbf{the above }$B_{z}%
$\textbf{ always goes to zero} on the surface (for any finite $z$), because of
its $1/\varrho$ - dependence (whereas for the 2nd gauge it is exactly zero on
the cylinder surface because of a delta function centered on the axis, see
Appendix A). Although $B_{z}$ is zero in the planar system, we see, however,
that the memory of a finite \textquotedblleft enclosed flux\textquotedblright%
\ in infinite 3D space remains, and it is this that actually causes the
proximity field influence. It is as if the cylinder axis has moved to infinity
in such a way that $B_{z}$ through the infinite space gives the same flux as
for the folded system, namely $B_{z}\rightarrow0$, but $\varrho\rightarrow
\infty$ in such a way that their product is either $-yB$ (dark area) or
$-d_{1}B$ (white area), which, in fact, \textit{are} the correct values of
$\mathbf{A}$ for our planar system, but now derived by a limiting procedure.
It is also interesting to emphasize that the nonlocal term of \cite{KM1,KM2}
for 2D static magnetic cases confirms (or, better, contains) this type of
proximity influence \textbf{directly in flat 2D space}, without the need of
any folding (or unfolding) or other limiting considerations (see Appendix B).

Regarding a connection of the above 2D ambiguity to, possibly, real physics,
note that, mathematically, a gauge transformation in the planar problem (upon
displacement of the origin $y_{0}$) \textit{is} an ordinary gauge
transformation (the gauge function is $\Lambda=-\Delta\Phi\frac{x}{L}$ (with
$\Delta\Phi$ being the change of flux that corresponds to the change of vector
potential $\Delta\mathbf{A}$) and is a smooth single-valued function
everywhere on the plane); when however we fold into a cylinder, the
corresponding $\Lambda$ turns out to be $\Lambda=-z_{0}BR\varphi$, that is
basically identical to the above planar $\Lambda$, but is now multiply-valued
(it has the usual discontinuity with respect to $\varphi$ appearing in all AB
phenomena in a cylindrically symmetric configuration). Hence in 3D the change
of origin $z_{0}$ is not unexpected to reflect real (difference in) physics
(similarly to the standard AB effect that introduces additional physics,
compared to a particle free of potentials); and, physically speaking, this has
to do with the \textit{different (infinite in number) arrangements} of the
total magnetic field (in 3D space) that \textit{all produce the same physical
arrangement on the side surface} (namely the same radial field component) and
therefore the same physical field-arrangement of our planar system. Note that
the formal appearance of $(z-z_{0})$ in $B_{z}$, actually reduces the
ambiguity to one with respect to \textit{displacements of the point where
}$B_{z}$ \textit{vanishes}. And that there is a great arbitrariness in placing
the point of vanishing $B_{z}$ somewhere in 3D space, although the 2D system
does not know of all this freedom -- it only senses the radial field, which is
\textit{always} (for \textit{any} of these $\mathbf{B}$-constructions) the
same $-$ in our case it is $B$ in the dark area and zero in the white. And
then, any such change of the location of the vanishing point $z_{0}$ on the
cylindrical surface involves relative displacements of the total $\mathbf{B}%
$-field in 3D space (relative to the cylinder), and this must be the source of
\textit{momentum transfer} to the particle. Indeed, such momentum transfer
(integrated in infinite time) turns out to be equal to $q\Delta\mathbf{A}/c$
(as can be shown by following lines of reasoning similar to those of i.e.
ref.\cite{Semon}) and gives a physical origin to the extra phases (of AB
type)\ picked up by the particle's wavefunctions upon change of $z_{0}$. The
crucial element is that our original planar system, with the pbc, is an
effectively compact system (since it can always be viewed as the
$R\rightarrow\infty$ limit of a compact cylinder), and due to the
compactification, the gauge transformations are not so innocent (they are
actually singular, and hide real physics), the nontrivial effects having as
origin the above noted momentum transfers. [Note also that, although in the
planar system $B_{z}$ \textit{vanishes everywhere}, the special point $z_{0}$
(or now $y_{0}$) has already been identified (as the unique point of local
vanishing of $B_{z}$ in the 3D companion system) \textit{before taking the
limit }$-$ something that will be of relevance in the experimental discussion later.]

To better appreciate the physics, we give in Appendix C a comparison to a
simpler system (with an in-plane electric field $E$ in the dark area). This
example demystifies somewhat this proximity effect and its gauge nature (now
\textit{momentum} reference levels being crucial, especially with respect to
the underlying $\Phi_{0}$-periodicity, compared to the usual (and
structureless) freedom associated with energy reference levels).

Effects of the above type are actually implicit in carbon
nanotubes\cite{Jakubsky13} (with the ambiguity now being reflected in a
quasiperiodicity parameter), and also have immediate applicability to planar
graphene (with no curvature)\cite{review}, although the above ambiguity has
not to our knowledge been discussed (or exploited) $-$ see however Appendix I
for our own suggestions on what to expect in such proximity measurements in
graphene and topological insulator surfaces.

\section{Generalizations}

\bigskip Let us briefly point out some consequences on previous works, and
discuss certain important generalizations, as well as issues of experimental
relevance (on how i.e. these proximity influences could be detected in the
laboratory)$\mathbf{:}$ \textbf{(i)} The above types of effects also appear in
connection with the concept of \textquotedblleft effective scalar
potential\textquotedblright\ that has been extensively used in previous works
(both on conventional systems\cite{PeetersMatulis} and on Dirac
materials\cite{PeetersMatulis2,PeetersMatulis3}) and in cases that the field
is \textit{accessible} to the particle (although this is not the focus of the
present work $-$ the case of \textquotedblleft forbidden\textquotedblright%
\ fields making our proximity effect more \textquotedblleft
striking\textquotedblright\ (or physically unexpected), see Appendix D). The
above noted gauge ambiguity applied here shows up as a
\textit{gauge-dependence of the effective scalar potential} (that seems to
have escaped notice in previous works, amounting to a large number of articles
by different groups), and it seems to affect even the \textit{qualitative
form} of this potential in the white area, bringing about important changes in
measurable quantities in either conventional or Dirac systems (briefly
discussed in Appendix D). \textbf{(ii)} The above folding procedure of our
dark-and-white system actually generalizes Laughlin's gauge argument on a
cylinder\cite{Laughlin}, where, however, the presence of the above $B_{z}$ is,
to our knowledge, rarely discussed. The addition of our white strip on the
surface of the usual Laughlin cylinder gives nontrivial consequences
\textbf{whenever the outside magnetic flux is not quantized }(see Appendix E,
on effective pumping and IQHE conditions induced from the outside).
\textbf{(iii) }One should note that all the phenomena predicted here should be
observable, independent of our (or any other) analysis of the $z_{0}%
$-ambiguity. One can give $z_{0}$ an absolute meaning (for a particular
\textit{cylindrical} system in the laboratory): it is the point in the 3D
folded system at which the $z$-component $B_{z}$ of the total 3D magnetic
field $\mathbf{B}$ (or its source, the current density $\mathbf{J}$) vanishes.
We can therefore determine this point $z_{0}$ in our 3D setup (see i.e. in
fig.3 of ref.\cite{Westerberg} the point where the magnetic lines are curved
in opposite directions), and then be careful to place our system of interest
(i.e. a strip with no field, exhibitting quantum coherence parallel to the
interface with the dark magnetic region) in a manner so that its basis (namely
the interface itself) is \textit{displaced} (by a small distance $d_{1}$) with
respect to $z_{0}$. Then, if this distance $d_{1}$ is such that the outside
magnetic flux is not quantized, then the above effects (a proximity influence
of this flux) should be present and measurable. (If they are not ever found,
then something is wrong with standard quantum theory and/or (classical)
electromagnetism). And, as shown earlier by a limiting procedure, these
proximity influences must survive even after the system becomes flat. However,
a question arises about cases when we \textit{start} with a \textit{strictly
flat} system, with no knowledge of the location of the $B_{z}$-vanishing point
of a corresponding 3D companion. For such cases, we will argue that we have
two options to consider$:$ for the 1st, see Appendix F (where it is shown that
a possibility still remains to have a nonlocal effect with no ambiguity), and
the 2nd is the case of actually having the $y_{0}$-ambiguity, which is now
physically unacceptable, and then\textbf{ }our criterion of proper behavior
(noted earlier) \textit{must be enforced}. This enforcement of elimination of
the ambiguity then seems to lead to \textbf{(a)} topological physics
(manifested as quantization of certain quantities, such as magnetic charge and
response functions), as well as to \textbf{(b)} connections and formal
analogies with other physics areas. Indeed, \textbf{(a) }recall that, in all
the above, essential use was made of the nonexistence of magnetic monopoles in
3D (the $\mathbf{\nabla\cdot B}=0$ law). But what if we had assumed that
magnetic monopoles exist? Our simplest finding on this (see Appendix
G)\textbf{ }is that imposition of our criterion of proper behavior (forced
elimination of the $z_{0}$-ambiguity) leads to quantization of fluxes external
to the \textquotedblleft white\textquotedblright\ system, so that, in the
limit that our white system shrinks to zero, the nonlocal term of
\cite{KM1,KM2} can serve as a \textit{probe of quantization} of the flux
through the magnetic region; and the enforced quantization of the nonlocal
term leads, in turn, to the quantization of magnetic charge according to the
Dirac condition\cite{Dirac}, and more generally, to the quantization of other
macroscopic quantities, that are related to quantized magnetoelectric effects
in an axion electrodynamical consideration\cite{Wilczek} (see Appendix H). In
particular, our criterion seems to nicely complement the recent proof of the
$2\pi$-periodicity of the axionic action\cite{VazifehFranz} by providing a
justification of the quantizations of certain separate 2D fluxes (one in space
and one in spacetime) that are crucial in the proof, justification that is not
given in ref.\cite{VazifehFranz} (see Appendix H for details). \textbf{(b)
}Apart from the above, there are much wider implications (mainly
\textit{physical}), but also relationships with other physics areas that one
can see \textit{formal} analogies with (see Appendix J on axions, $\theta
$-vacuum sectors\cite{Asorey,Zhitnitsky}, Gribov copies\cite{Gribov}, but also
connections with certain open problems in mechanics\cite{BerryShukla} and in
thermodynamics\cite{Kiehn}), that certainly necessitate further investigation
of an interdisciplinary character.

\section{Discussion}

Even however the simplest outcome of the present theoretical work $-$ that it
is in principle \textit{possible} to have effects without fields, \textit{in
the simply-connected plane}, that originate from outside of our system and
that affect its physical properties $-$ is remarkable, and if true, extremely
important in experimental work on fundamental physics as well as in practical
applications. First, the most obvious use is for an easier experimental
detection of AB effects: these can be indirectly measured in a
simply-connected system and without enclosed and confined fluxes in the
laboratory $-$ hence with considerably lesser problems of leakage of magnetic
lines, compared to the enclosed confined configurations typically used. Then,
the already noted possibility of violation of Bloch theorem (especially if our
\textquotedblleft white\textquotedblright\ (no-field) system is periodic along
the interface direction) is worth emphasizing. The violation is due to the
presence (on the system) of the extra vector potential (from proximity with
the outside $B$-field), hence due to the hidden AB effect caused by the 3D
companion system, and it leads to AB-type of modifications of translation
operators etc. that are used in the standard proof of the Bloch theorem. Note
that these modifications are not the same as the well-known modifications of
Bloch theorem in an IQHE system (such as the ones studied i.e. in
\cite{Kohmoto}) with the particle being inside a field $-$ in our case we
always have $B=0$ on the particle. We therefore eventually expect
modifications in the form of wavefunctions (essentially of an AB-type), and
these will now be different from the standard Bloch forms; in such a case, one
can first gauge away the \textquotedblleft proximity-induced $\mathbf{A}%
$\textquotedblright, with the consequence of the extra appearance in the
boundary conditions of a crystal momentum (parallel to the strip) that is
essentially the kinematic momentum. And then, by adiabatically changing the
special point $y_{0}$ in a direction transverse to the strip by a
\textquotedblleft cycle\textquotedblright\ (meaning that the corresponding
change of flux is equal to $\Phi_{0}$, see Appendix E), we can have the
crystal momentum moving from one edge to the other of the (parallel) Brillouin
zone, and hence induce new effects (or transitions, in a one-electron picture
$-$ not to mention the possible novel effects in the presence of
electron-electron interactions, such as formation of composite fermions
\textit{at zero magnetic field}\cite{Gvozdikov}). It is also interesting, and
potentially useful experimentally, that, in cases when both electrons and
holes are considered, the Berry phase picked up during such a cycle seems to
contain not only an AB part (as derived by Berry in the transported rigid box
around an AB flux\cite{Berry}), but also a term directly related to the
\textit{electric current}, similarly to what happens in an AB
ring\cite{martha,kyriakosring}. Finally, a periodic (or even quasiperiodic,
i.e. Fibonacci) arrangement of magnetic strips (on a cylinder, or in the plane
with pbc parallel to the strips), each one containing a rational flux
$f\Phi_{0}$ (with $f=\frac{p}{q}$, $p,q$ integers), would be an interesting
system to consider, with new (in)commensurability effects expected (not of the
Hofstadter type\cite{Hofstadter} where we have a nonzero $B$-field), that will
be a result of the interplay between the gauge nonlocality of this work and
the (quasi)periodicity of the structure $-$ behavioral patterns that will be
possibly useful for novel applications in intelligent devices.

\section{Conclusions}

Regarding all the above, it is for the experiment to give the verdict, but it
is fair to say that we have provided in this work strong theoretical evidence
(in fact a rigorous proof) that the physics of a system can partly be dictated
not only by local physical laws but also by nonlocal influences (from remote
regions in spacetime) with a gauge character. We reemphasize that, because of
this, it is in principle \textit{possible} to have effects without fields,
\textit{in the simply-connected plane}, that originate from outside of our
system and that affect its physical properties $-$ something remarkable, and
important at least for novel applications. And although we have focused on
orbital physics, there are well-defined steps (through boosts to properly
moving frames) that lead to spin-physics as well (see Appendix J, on how a
hidden Aharonov-Casher effect\cite{AC} is also expected to be involved). Of
course one can simply take spin into account in all the above problems in a
direct formal manner, or in a similar fashion to calculations that have been
carried out in graphene or in other Dirac systems, when these are in AB
configurations (see i.e. Appendix I) $-$ although a generalization of the U(1)
gauge character of the nonlocal effects proposed here to cases with a
spin-orbit coupling (now with an SU(2) character) would have an additional
importance for modern applications and, as already noted, deserves a separate
note. This arguably demonstrates that, if the above proximity effects turn out
to be real, the experimental and application possibilities of exploiting them,
as well as their generalizations, seem to be almost endless.

\bigskip

\bigskip\ \ \ \ \ \ \ \ \ \ \ \ \ \ \ \ \ \ \ \ \ \ \ \ \ \ \ \ \ \ \ \ \ \ \ \ \ \ \ \ \ 

\ \ \ \ \ \ \ \ \ \ \ \ \ \ \ \ \ \ \ \ \ \ \ \ \ \ \ \ \ \ \ \ \ \ \ \ \ \ \ \ 

\ \ \ \ \ \ \ \ \ \ \ \ \ \ \ \ \ \ \ \ \ \ \ \ \ \ \ \ \ \ \ \ \ \ \ 

\bigskip

\section{\bigskip APPENDICES}

\subsection{Cylindrical geometry and flux-cancellations}

Recall that $\left(  \mathbf{\nabla\times A}\right)  _{\varrho}=\frac
{1}{\varrho}\left(  \frac{\partial A_{z}}{\partial\varphi}-\frac
{\partial\left(  \varrho A_{\varphi}\right)  }{\partial z}\right)  $, $\left(
\mathbf{\nabla\times A}\right)  _{\varphi}=\left(  \frac{\partial A_{\varrho}%
}{\partial z}-\frac{\partial A_{z}}{\partial\varrho}\right)  $ and $\left(
\mathbf{\nabla\times A}\right)  _{z}=\frac{1}{\varrho}\left(  \frac
{\partial\left(  \varrho A_{\varphi}\right)  }{\partial\varrho}-\frac{\partial
A_{\varrho}}{\partial\varphi}\right)  $. For our gauge $\mathbf{A=(}%
0,-zB\mathbf{,}0\mathbf{)}$ in the dark area and $\mathbf{A=(}0,-d_{1}%
B\mathbf{,}0\mathbf{)}$ in the white area, these lead to: $B_{\varrho}%
=\frac{1}{\varrho}\left(  -\frac{\partial}{\partial z}\left(  -\varrho
zB\right)  \right)  =B$ in the dark area, and $B_{\varrho}=\frac{1}{\varrho
}\left(  -\frac{\partial}{\partial z}\left(  -\varrho d_{1}B\right)  \right)
=0$ in the white area as required; we also obtain $B_{\varphi}=0$ in both
areas, and finally $B_{z}=\frac{1}{\varrho}\left(  \frac{\partial}%
{\partial\varrho}\left(  -\varrho zB\right)  \right)  =-\frac{zB}{\varrho}$ in
the dark area, and $B_{z}=\frac{1}{\varrho}\left(  \frac{\partial}%
{\partial\varrho}\left(  -\varrho d_{1}B\right)  \right)  =-\frac{d_{1}%
B}{\varrho}$ in the white area. To make the cancellations of the text easily
visible, take the special choice $z_{1}=0$ (the floor of the cylinder being at
the origin (at the floor of the dark strip)) and for $z_{2}=z$ (the ceiling of
the cylinder, lying either (a) inside the dark or (b) inside the white area);
the curved side-surface then consists of either (a) just a lower part of the
dark strip or (b) the entire dark area (a full curved strip, going around the
axis and always lying on the curved cylindrical surface) together with a lower
part of the white folded strip. Then indeed, the flux of $B_{z}$ through the
ceiling is $\int\int B_{z}\varrho d\varrho d\varphi=-zB2\pi R$ if $z$ is
inside the dark area, or $-d_{1}B2\pi R$ (hence a constant) if $z$ lies inside
the white area; and we see that, in either case, it indeed cancels out the
radial $B$-flux (which is $zB2\pi R$ in the dark area and the constant
$d_{1}B2\pi R$ in the white area, either of which can de determined by use of
the proper $B_{\varrho}$ as given above); and the flux of $B_{z}$\ through the
ceiling is also identical to the value of a closed integral of the
corresponding $\mathbf{A}$ around the cylinder (which is $\int A_{\varphi
}Rd\varphi=-zB2\pi R$ or $-d_{1}B2\pi R$) as expected, this way clearly
demonstrating that the above cancellation is actually due to the standard AB
effect (due to the $B_{z}$-flux that \textit{is enclosed} by the particle's region).

In the 2nd gauge discussed in the main text, namely $\mathbf{A=(}0,-\frac
{zBR}{\rho}\mathbf{,}0\mathbf{)}$ in the dark area and $\mathbf{A=(}%
0,-\frac{d_{1}BR}{\rho}\mathbf{,}0\mathbf{)}$ in the white area, from
application of the above we see that the radial magnetic field is not a
constant $B$ in all space (for $z\leq d_{1}$) as before, but it is now
$B_{\varrho}=\frac{BR}{\varrho}$; we see therefore that we also have
flux-cancellations that occur radially, namely through any internal
cylindrical and any external cylindrical surface as expected. It also naively
seems that $B_{z}$ goes away (since $\varrho A_{\varphi}$ is now independent
of $\varrho$), which however is only true for $\varrho\neq0$; if $\varrho=0$
is included, $B_{z}$ actually becomes proportional to a 2-dimensional delta
function at $\varrho=0$\textbf{:} it is well-known in the AB literature (and
is a crucial result in the Chern-Simons transformation in many-body physics
that leads to composite fermions) that $\nabla\times\frac{1}{\varrho}\hat
{e}_{\varphi}=2\pi\delta^{(2)}(\mathbf{\varrho})\hat{e}_{z}$, so that the
\textit{curl} of the above $\mathbf{A}$ has (in the dark area) a $z$-component
equal to $-2\pi RzB\delta^{(2)}(\mathbf{\varrho})$, giving again a return flux
along $z$-axis for this choice as well, that will cause similar cancellations
as above (i.e. fluxes in the $z$ and radial directions will again cancel out).
So, overall, we have the same physical interpretation as in the 1st gauge.

\subsection{Nonlocal Terms}

An immediate first understanding of the proximity influence of adjacent fields
(that will also be practically useful later (as a probe, or detector of
quantizations)) comes from a recent theory\cite{KM1,KM2}, that leads to cases
where the \textquotedblleft gauge function\textquotedblright\ $\Lambda$
(defined by $\mathbf{A}=\mathbf{\nabla}\Lambda$ at the point of observation
$(x,y)$) does not only contain the standard integrals over potentials, but can
also contain nonlocal terms of remote fields (this occurring whenever the
paths of integrals pass through these fields $-$ the point of observation
being however always outside them). This is exactly what we are witnessing in
the first example of the main text (and in the simplest possible static
magnetic case). Indeed, if we use the 2nd solution of ref.\cite{KM1} for
$\Lambda$ (see eq.(9)), where we have vanishing integrals over paths, then we
find a nonzero nonlocal term, namely $\Lambda\backsim\int_{y_{0}}^{y}%
\int_{x_{0}}^{x}\mathbf{B(}x^{\prime},y^{\prime})dx^{\prime}dy^{\prime}$
\ that must be \textit{independent of} $y$, as required by an attached
condition (constraint) to this particular solution (see condition in eq.(9) of
ref.\cite{KM1}). This 2nd solution \textit{is}, for the particular case of our
horizontal strip, \textit{indeed independent of} $y$ [if we move the point of
observation $(x,y)$ up and down, the flux enclosed inside the
\textquotedblleft observation rectangle\textquotedblright\ does not change],
hence it is \textit{acceptable}; it actually turns out to be $-\Phi
_{dark}\frac{x}{L}$, and this also yields the same result as in the main text
(namely $\Lambda=-\Phi_{dark}\frac{x}{L}=A_{_{0}}x,$ and the rest can be
worked out as before, leading again to eq.(1) of the main text). As seen from
the above, the origin-dependence is built in the form of the nonlocal terms
(so they are expected to offer a natural language to describe these proximity
influences and the possible gauge ambiguities $-$ see later sections on how
quantizations emerge from essentially this). Also note that, in the main text,
the nonlocal terms were shown to contain (or to have knowledge of) the hidden
AB influences in higher spatial dimensionality, without the need of folding or
unfolding or other limiting considerations (see main text).

The exact results of the theory of nonlocal terms of refs.\cite{KM1,KM2} have
recently been derived starting from a rather surprising advance in elementary
calculus\cite{Kyriakos} $-$ based on local expansions in 2D around the
observation point, combined with further geometric reasoning in properly
applying Stokes' theorem $-$ and apparently they have been overlooked in the
physical but also in the mathematical literature (as generalized solutions of
$\nabla\Lambda=\mathbf{A}$, on the simple-connected plane, \textit{when
}$\mathbf{\Lambda}$\textit{ is not defined as a decent function everywhere} on
the plane (i.e. due to the presence of remote fields)); it has also been
proven\cite{KM} that these nonlocal forms $-$ highly nonlinear in the
potentials $-$ are also Lorentz invariant, nicely generalizing therefore the
usual 4-vector form of the standard electromagnetic Lagrangian (in its
interaction with matter fields) that is only linear in the potentials.

\bigskip In the discussion in Appendix E later, on possible \textquotedblleft
proximity devices\textquotedblright, it will be noted that one can use other
more sophisticated types of procedures (of inducing topological phenomena from
outside the system) based on nonlocal terms in \cite{KM1,KM2} that involve
$t$-dependent electric fields and electric scalar potentials. In this context
it should be reminded that such $t$-dependent nonlocal terms were shown in
ref.\cite{KM1,KM2} (through examples with involvement of electric fields) to
indirectly \textbf{protect relativistic causality} by leading to cancellations
of causality-violating terms (that seem to be silently accepted in current
theories). Hence one may expect in the remote field-influence on such devices
some type of \textquotedblleft causal indeterminism\textquotedblright\ (with
causality \textquotedblleft hidden\textquotedblright\ in the phases of the
quantum wavefunctions (by exploiting the Lorentz invariance of the
time-dependent nonlocal terms, that we noted above)), and this actually
occurring without the involvement of uncertainty principle (that currently
seems to be considered absolutely necessary, to protect from
causality-violation\cite{Aharonov}). Summarizing, it seems that, generally
speaking, in a number of different ways one can induce conditions of, at least
topological (quantized) pumping of some quantity, resulting from manipulations
from outside of our system, and in ways that are expected to respect
relativistic causality. [See also how the nonlocal term-solutions can be
related to irreversibility in thermodynamics and to recent open problems in
mechanics, in Appendix J.]

\subsection{Comparison with a simpler problem}

To better appreciate the physics of our first example in the main text, let us
make a physical comparison with a different (and simpler, or more familiar)
system$\mathbf{:}$ if, instead of a $B$ across the dark region, we had chosen
an in-plane electric field $\mathbf{E}$ (static and homogeneous, with i.e.
direction pointing downwards), then \textit{there would still be a gauge
}(\textit{or origin-dependent})\textit{ ambiguity }(\textit{now for the
\textbf{scalar potential }}$V$),\textit{ attributed to a freedom of choice of
\textbf{energy}-reference level}. Indeed, inside the dark region we would now
have a linear scalar potential increasing upwards, so that, if we chose again
to set the zero of potential at our origin (hence to ground the bottom of the
dark region), then the bottom of the white area (and from that point upwards,
the entire white area) would have a nonzero constant and uniform scalar
potential (equal to $+Ed_{1}$), since the whole white area, where our particle
resides, is an equipotential region -- like a region outside the positive
plate of an ideal plane capacitor whose negative plate is grounded. This
constant potential (and through this the \textit{outside} (adjacent) field
$E$) will contribute \textbf{additively} to the particle energies; and
although this additivity is physically obvious, one can also formally see it
through the use of gauge transformations in full generality, see below -- note
in particular that, if the electric field did not last for ever (i.e. if it
were generally $t$-dependent), physics would be affected differently. And if
we now displace the origin (hence the zero-level of scalar potential), the
value of potential inside the white area $-$ and hence the energies of the
particle $-$ will change accordingly (the \textquotedblleft priviledged
case\textquotedblright\ being to ground the ceiling (rather than the floor) of
the dark area, so that we have no effect at all). In any case, it seems that
even in this trivial problem, we can have an influence of an \textit{outside}
(adjacent) field, and also an \textit{ambiguity} of this influence with
respect to the placement of some origin: there is an infinite freedom of
choice of this origin that gives (additively) different energies to our
particle. But, again, such changes (of gauge) would be considered as innocent
(and natural), namely as mere changes of energy-reference level. In our case
of outside $B$ and the involvement of vector potentials (as in the examples of
the main text), we actually witness a similar ambiguity (it is again a freedom
of choice of a reference level), but it now concerns \textbf{momenta (}rather
than energies). And this, we claim, introduces useful and interesting physics,
basically because of the manner in which the vector potential couples to the
Hamiltonian (the minimal coupling) and also because of the \textit{structure}
underlying the freedom of choice of momentum reference levels (namely, the
\textbf{periodicity}, with period $\Phi_{0}$) $-$ a property that the (simply
additive) energies \textit{do not} have.

Finally, concerning a brief argument in order to see the \textit{formal}
difference between the two cases compared above (one with static magnetic and
the other with static electric field in the dark region), note that the
nonlocal term in general time-dependent cases of ref.\cite{KM1,KM2} for an
arbitrary electric field $\mathbf{E}$ has the form $\Lambda(y,t)=-c\int
_{t_{0}}^{t}dt^{\prime}\int_{y_{0}}^{y}dy^{\prime}E(y^{\prime},t^{\prime})$
which, in our case of homogeneous and static $\mathbf{E=}E\hat{e}_{y}$ and
with grounded the basis of the dark strip, is simply $-cV$, with $V$ the value
of the scalar potential on the ceiling (and inside the entire white area,
being $+Ed_{1}$) $-$ we see therefore that the \textbf{outside} (and
inaccessible) electric field contributes to the white area even in this
trivial situation. Indeed, the above $\Lambda$ leads to time-dependent
wavefunctions in the $y$-direction that will now be of the form $\Psi
(y,t)\thicksim e^{i\frac{qV}{\hbar}(t-t_{0})}e^{-i\frac{\epsilon}{\hbar
}(t-t_{0})}\psi(y)$, which, after separation of the $t$-variable will lead to
a static Schr\"{o}dinger equation for $\psi(y)$ that will now have as a
parameter the combination $(\epsilon-qV)$ rather than just the energy
parameter $\epsilon$, leading at the end to energies of our particle that are
$\epsilon(V)=\epsilon(V=0)+qEd_{1}$, namely just an additive contribution to
the energies (due to the scalar potential on our system, or, note, due to the
\textbf{outside} field) as physically expected. (Note again that, if the
electric field were $t$-dependent, things would not be so trivial). However,
as already noted, our case of an outside \textit{magnetic} field is much more
interesting: energies at the end are now affected not additively, but
indirectly through the changes of momenta (because of the minimal coupling of
the vector potential in the hamiltonian) and with the $\Phi_{0}$-periodicity
being essential, as discussed in the main text.

\subsection{Accessible fields}

Here is a brief discussion on proximity cases with fields that are
\textit{accessible} to the particle (which are of course easier to achieve
experimentally $-$ although such cases of particles entering the adjacent
fields may take away the \textquotedblleft mystery\textquotedblright\ of the
proximity field-influence, and is not quite the focus of the present work).
However, it is important to point out that systems with magnetic strips or
barriers have been often discussed in the literature (for accessible fields)
by matching methods, through the use of the effective scalar potential noted
in the main text, and this has been done for both conventional systems and
Dirac materials. First, in a setup such as the text's orthogonal strip with a
conventional parabolic hamiltonian, the effective scalar potential (which is
$k_{x}$-dependent) contains $A_{x}(y)$, and this makes it gauge-dependent,
something that also seems to have escaped notice. It is important to emphasize
again that the changes of gauge do not cause changes of a purely additive
energy type, but the physics is dramatically affected through the qualitative
change of the \textit{form} of the effective scalar potential in the outside
\textquotedblleft white\textquotedblright\ area (see i.e. fig. 1(b) of the
first of ref.\cite{PeetersMatulis}), this form depending on the combination of
$d_{1}$ and the sign of $k_{x}$ (see below). Indeed, it turns out that
$V_{eff}(y)=\frac{\left(  \hbar k_{x}-\frac{q}{c}A_{x}(y)\right)  ^{2}}{2m}$,
and in the white area $A_{x}(y)$ is a constant $-d_{1}B$, whose value is
$d_{1}$-dependent, and it is matched with the form of $V_{eff}$ as this comes
from inside the field at the interface; inside the field we have
$V_{eff}(y)=\frac{1}{2}m\omega_{c}^{2}(y-y_{0})^{2}$ with $y_{0}=-k_{x}%
\frac{\hbar c}{qB}$, and it is clear that if $d_{1}$ is not an integral
multiple of $y_{0}$, then we have nontrivial consequences on the form of the
potential (and therefore of the solutions) outside (whereas if $d_{1}=Ny_{0}$,
with $N$ integer, then $\Phi_{dark}$ is quantized and there is no new effect).
In fact, the energy spectrum comes out as $\epsilon_{n_{x},n_{y}}=\frac
{\hbar^{2}}{2m}\frac{\pi^{2}n_{y}^{2}}{d^{2}}+\frac{\hbar^{2}}{2m}\left(
\frac{2\pi}{L}\right)  ^{2}(n_{x}-\nu)^{2}$, with $\nu=\frac{qdB}{hc/L}%
=\frac{\Phi_{dark}}{\Phi_{0}}$, and if $\nu$ is not an integer the spectrum is
not equivalent to the $\nu=0$ case.

In the case of Dirac materials, by using the Dirac Hamiltonian $H=v_{f}%
\mathbf{\sigma\cdot\Pi}$ (with $\mathbf{\Pi=p-}\frac{q}{c}\mathbf{A}$ the
kinematic momentum) and with ansatz $\Psi_{i}(x,y)\thicksim\Psi_{i}%
(y)e^{ik_{x}x}$ (with $i=1,2$ denoting the components of a spinor) it turns
out that for the white area we have to solve a system of Schr\"{o}dinger-like
equations, namely $\left(  -\hbar^{2}\frac{\partial^{2}}{\partial y^{2}%
}+\left(  \hbar k_{x}-\frac{q}{c}A_{x}(y)\right)  ^{2}\right)  \Psi
_{1,2}(x,y)=\frac{E^{2}}{v_{f}^{2}}\Psi_{1,2}(x,y)$, and we clearly see a
similar effect as in the nonrelativistic system (the detailed solution will be
given elsewhere). In case that the dark strip has no integrally-quantized
flux, the solution is again not equivalent to the case $\nu=0$, especially so
for a disk-geometry (with accessible $B$). Once again, at the bottom of this
is phase-physics (and the phase-mismatch around the cylinder (or around the
center in the disk) when $\Phi_{dark}$ is not quantized) [details to be
published\cite{Giwrgos}]. And if we follow this method of effective scalar
potential for our original striped system with the magnetic region being again
\textit{inaccessible}, then it turns out (in a quite different manner from
what we did in the main text) that the energy spectrum in the white area is
identical to eq.(1) of the text, with $f=\nu=\frac{qd_{1}B}{ch/L}$ which is
$\Phi_{dark}/\Phi_{0}$, in agreement with our gauge transformation mapping
technique. Hence the use of the effective scalar potential method $-$ and the
solution based on matching conditions in a direction \textit{transverse} to
the interface $-$ seems to lead to the same results as those of a
phase-mismatch analysis \textit{parallel} to the interface.

In a similar vein, systems such as Zygelman's recent work\cite{Zygelman} are
also expected to be affected $-$ if we impose periodic boundary conditions
parallel to the strip $-$ whenever the flux of the strip is \textit{not
quantized} (and it is easier to see this if we take the strip to be a delta
function). A detailed solution will be given elsewhere\cite{Giwrgos} with the
direct use of the concept of pseudomomentum (and how it is affected across the
interface from inside to outside the field) and also of a new concept of
pseudo-angular-momentum\cite{Kyriakos2} for the corresponding problem in a
disk geometry. However, note again that the focus of the present work is
\textit{not} on fields sensed by the particles, but in \textquotedblleft
forbidden\textquotedblright\ fields, because it is these cases that may make
the effect of nearby fields more striking (or physically unexpected).

\subsection{Generalizing Laughlin's argument}

The folding procedure of our dark-and-white system into a cylinder, used in
the main text, actually generalizes Laughlin's gauge argument on a
cylinder\cite{Laughlin}, where, however, the presence of the crucial $B_{z}$
is, to our knowledge, rarely discussed. In the standard Laughlin's argument,
with a radial $B$ being \textit{everywhere} on the cylinder's curved
cylindrical surface, one can actually understand the well-known translational
symmetry breaking\cite{JansenLiebSeiler} $-$ where the equilibrium positions
of the standard Landau wavefunctions ($y_{0}=k_{x}l^{2}$ in planar language,
with $l=\sqrt{\hbar c/eB}$ the magnetic length) become
priviledged\cite{JansenLiebSeiler} $-$ by the special consideration of this
additional $B_{z}$ created due to folding$\mathbf{:}$ as we saw, the AB flux
enclosed by a horizontal circle (lying on the cylindrical surface) around the
axis depends on the \textquotedblleft height\textquotedblright\ $z$ (due to
the presence of the $B_{z}=-\frac{zB}{\varrho}$), so that, if we want
immediate wavefunction single-valuedness around the cylinder, we indeed need
special $z$'s so that the enclosed AB flux (at that height) is quantized (in
integral multiples of $\Phi_{0}$). It is straightforward to see that this
requirement gives immediately the priviledged $z_{0}$'s (or equivalently the
above equilibrium positions $y_{0}$'s for the standard flat Landau problem in
the Landau gauge). But in our generalized system, with the area of interest
(where the particle resides) being only a white strip on the cylindrical
surface (with no field $B$ inside it), one finds that there are nontrivial
consequences (due to remote field influence) on this white area,
\textbf{whenever the outside magnetic flux is not quantized}. This we saw with
inaccessible fields, but it seems to also occur for accessible ones as well
(see Appendix D). In such case of non-quantized $\Phi_{dark}$, the
wavefunction single-valuedness (or periodic boundary conditions along $x$) in
the white area is \textit{not} automatically satisfied, and it is its
enforcement that leads to a modification of physical properties, hence to the
remote influence of the adjacent magnetic field that we saw. A plausible
question would then be: is there a \textquotedblleft remote (or proximity)
influence of the IQHE type\textquotedblright\ that might affect the particle,
although this resides \textbf{outside} the field $B$ (hence, equivalently, a
quantum Hall effect in zero-field)? There is a great deal that can be said on
this, especially in cases that $B$ is accessible to the particle $-$ i.e. in
relation to \textquotedblleft magnetic edge states\textquotedblright\ in the
interface\cite{Hislop}, snake states\cite{Mueller} etc. to be discussed in a
more focused paper (the main conclusion for now being that we must have
nontrivial dissipationless edge currents in the interface that, in any case,
are expected, as the persistent currents associated with the hidden AB effect,
being therefore proportional to $\partial\epsilon/\partial y_{0}$); but even
without details, we will point out as certainly true that one can generate (or
simulate) IQHE conditions on our system (always a \textquotedblleft white
area\textquotedblright, with no $B$) with a pulsed outside electric field $-$
rather than the static field case discussed in the main text and in Appendix A
$-$ which, due to its time-dependence, can induce IQHE type of effects inside
our field-free system (a case involving remote electric fluxes in spacetime).
An even simpler way is the main text's original example of a magnetic field
$B$ in the dark area, which however is not static but slowly (adiabatically)
changing with time, or, alternatively, a fixed $B$ while our origin $y_{0}$ is
being displaced slowly (and transversely to the interface) between two values
\textit{that correspond to a change of flux in the dark area equal to }%
$\Phi_{0}$ (this would then define a \textquotedblleft cycle\textquotedblright%
). This way one can achieve charge pumping (with slow variation of $B$ or of
$y_{0}$ or proper combination of both) as in the case of Laughlin's
cylinder\cite{Laughlin} (replacing the much harder to build externally applied
varying \textit{enclosed} AB flux). After a cycle, there must be an integer
number of electrons transported from one side of the system to the other
(along the $y$-direction), a well-known topological quantum effect due to
Thouless\cite{Thouless}. Or one can use other more sophisticated types of
procedures based on nonlocal terms in refs \cite{KM1,KM2} involving general
$t$-dependent electric fields and electric scalar potentials. Summarizing, it
seems that, in a number of different ways, one can induce conditions of, at
least, topological (quantized) pumping of some quantity, resulting from
manipulations from \textit{outside} of our system, and, in fact, in ways that
are expected to \textit{respect relativistic causality}, as noted in the main
text (and clarified in Appendix B).

\subsection{How to measure the nonlocality in a strictly planar system}

First, for a cylindrical arrangement, we have seen in the main text that the
special vanishing-$B_{z}$ point ($z_{0}$) is unique and identifiable, and
survives in the $R\rightarrow\infty$ limit, so that the remote influences that
are the focus of the present work \textit{must survive even after the system
becomes flat}; and although in the completely planar system $B_{z}$
\textit{vanishes everywhere}, we have already identified the absolute
reference point $z_{0}$ (or $y_{0}$) \textit{before the limit }(as the unique
point of vanishing $B_{z}$ that existed in the companion 3D system). Hence, by
using this $y_{0}$, we can achieve (or measure) all the types of proximity
effects discussed in the main text in the same way (namely, by placing our
white area in a properly displaced manner with respect to this $y_{0}$).

However, for strictly planar system, when we have no knowledge of the $B_{z}%
$-vanishing point of a corresponding 3D companion, we argued in the main text
that we have two options to consider, and here we focus on the first$:$ If we
have a large-width ($d_{1}$) magnetic area, it is quite possible that,
generically, this would behave as if it were produced by a corresponding long
cylinder (in the usual theoretical limit $R\rightarrow\infty$) with its
special vanishing-$B_{z}$ point ($z_{0}$) being in the \textit{middle} of its
\textit{finite} length (see again fig.3 of ref.\cite{Westerberg} for such a
system); this is for symmetry reasons and due to the fact that all expressions
of the fields used here (and in fact in the whole literature) are actually
exact only in the case of \textit{infinite} cylinders -- the middle of a long
cylinder being therefore slightly preferred (as being the point that is more
distant from both cylinder-ends, and also because, due to its symmetrical
placement, it is a better representative of the infinite-cylinder theory). If
this turns out to be correct, then this suggests an obvious experimental way
on how to place our \textquotedblleft white\textquotedblright\ area: $y_{0}$
can be taken to be in the middle of the width of the flat dark area, and then
our white system must be placed as described in the main text. (Note that in
this case it will be the \textit{half} of the total outside flux that will
remotely influence our white area, and one has interesting possibilities to
study (even in case that the total outside flux is $\Phi_{0}$), if e.g. the
spin of the electron is included). In fact, a slightly better experimental
suggestion would be to have \textit{two} systems of interest
(\textquotedblleft white areas\textquotedblright, i.e. they could be identical
graphene samples), separated by the above (inaccessible) wide magnetic region,
and then make measurements (i.e. of persistent current) in one system
\textit{or} the other; the point is that, no matter where $y_{0}$ is located,
\textit{at least one} of the two systems \textit{must} be affected by
proximity (if i.e. it happens that $y_{0}$ is at the edge of one area, giving
no effect on this system, then the same $y_{0}$ is necessarily displaced with
respect to the 2nd system; so proximity influence on the 2nd system is
guaranteed, if the intermediate flux is not quantized, and we can measure
nontrivial effects in this 2nd system $-$ and it is interesting to note that,
if $y_{0}$ is indeed in the middle of the magnetic region, as we hoped
earlier, then now, in the present setup, both systems will be affected
equally). If all this does not work (meaning that there is no memory of
a\textit{ unique} $y_{0}$, a remnant of the theoretical limit), then the lack
of knowledge of a 3D companion is indeed complete, or equivalently this gives
rise to the earlier discussed ambiguity. In such case, as noted in the main
text, our criterion of proper behavior \textit{must be imposed} (see the
Appendices that follow).

\subsection{Dirac monopoles}

We are here interested in cases where (effective) magnetic monopoles are
present. Note that, already in the case of the \textquotedblleft Laughlin
cylinder\textquotedblright\ of the main text $-$ with the usual in the
literature practice of not any mention of the extra $B_{z}$ that originates
from folding of the original flat system $-$ it is seen that the radial $B$ in
3D space \textit{must} be a result of a linear magnetic monopole distribution
(along the $z$-axis) $-$ since a purely radial field violates the
$\mathbf{\nabla\cdot B}=0$ law (as there is a nonzero net flux outwards and,
therefore, magnetic monopoles must be invoked to justify it). And starting
with an additionally placed extra narrow ($d\thicksim0$) white strip (with no
field) that goes around the axis on the cylindrical surface, and imposing our
criterion of decency (elimination of the gauge ambiguity) in the limit
$d\rightarrow0$ one obtains the well-known quantization of the $B$-flux in the
dark area, and from this it comes out that the monopole charge must also be
quantized. It is however easier to see this with a similar argument in Dirac's
spherical geometry, with a single magnetic monopole at the center of a
sphere\cite{Dirac}; enforcement then of our criterion on a small white
circular section on the spherical surface around, say, the north pole, and in
the limit that this section shrinks to zero (to the north pole), gives that
$\Phi_{sphere}=N\Phi_{0}$, $N$ integer, which in turn leads to the well-known
Dirac's quantization condition for the magnetic charge density $\rho_{m}$,
namely $\rho_{m}=e_{m}\delta(\mathbf{r})=\frac{Nhc}{4\pi e}\delta(\mathbf{r}%
)$, consistent with the quantization of $\Phi_{sphere}$ (to check it, recall
that the radial field created by the monopole is $\mathbf{B=}\frac{e_{m}}{4\pi
r^{2}}\hat{e}_{r}$). [The small white section, with no field, can be achieved
through the use of 2 identical spheres that are tangential to each other at a
point (which will become the above north pole), with equal magnetic charges on
each center -- since at the tangential point the two separate fields are
opposite, and they will cancel out to yield a zero total field as required for
the argument; although one must be even more careful for the shake of rigor
[i.e. it actually turns out that we need 3 spheres, because apart from the
point of observation $\mathbf{r}$, we also want the initial point of
integration $\mathbf{r}_{0}$ (that shows up in the expression of nonlocal
terms) to lie outside $\mathbf{B}$, so that we actually need to consider 3
tangential spheres of equal radius with their centers lying on the $z$-axis,
and with each one having a magnetic monopole at its center (of equal magnetic
charge each); $\mathbf{r}$ will then finally be at the north pole and
$\mathbf{r}_{0}$ at the south pole of the \textit{middle} sphere (with both
points being at zero total field, because of the cancellations), for which
middle sphere we can then apply our above argument.]]

We see therefore that, by formally enforcing the elimination of this gauge
ambiguity in a closed system, the nonlocal term can indeed play the role of a
probe of (or a detector of) quantization of macroscopic quantities (although,
it should be noted, \textit{we are merely at the level of wavefunction
phases}). A plausible question then is: can such a type of argumentation be
followed to other more complicated cases? We answer positively and we give
below, in Appendix H, some considerably more sophisticated examples, by
considering topologically nontrivial systems, which $-$ as has been shown
recently\cite{axion,axion2} $-$ seem to need axion electrodynamics to describe
their magnetoelectric response properties.

\subsection{Axions}

Let us first recall axion electrodynamics (but with inclusion of magnetic
monopole terms, since they will be useful in our discussion of the Witten
effect\cite{Witten} further below). Axion electrodynamics can be described by
the augmented Maxwell's equations\cite{Wilczek}

$\mathbf{\nabla\cdot E=}4\pi\left(  \rho+\rho_{\theta}\right)  $

$\mathbf{\nabla\cdot B=}4\pi\rho_{m}$

\bigskip

$\mathbf{\nabla\times E=}-\frac{1}{c}\frac{\partial\mathbf{B}}{\partial
t}-\frac{4\pi}{c}\mathbf{J}_{m}$

\bigskip

$\mathbf{\nabla\times B=}\frac{1}{c}\frac{\partial\mathbf{E}}{\partial
t}+\frac{4\pi}{c}\left(  \mathbf{J}+\mathbf{J}_{\theta}\right)  $

where the extra axionic charge and current densities are defined by
($\alpha=e^{2}/\hbar c$ is the fine structure constant)

$\rho_{\theta}=-\frac{\alpha}{\left(  2\pi\right)  ^{2}}\mathbf{\nabla\cdot
}\left(  \theta\mathbf{B}\right)  =-\frac{\alpha}{\left(  2\pi\right)  ^{2}%
}\left(  \mathbf{\nabla}\theta\mathbf{\cdot B+}\theta\mathbf{\nabla\cdot
B}\right)  $

$\mathbf{J}_{\theta}=\frac{c\alpha}{\left(  2\pi\right)  ^{2}}\mathbf{\nabla
\times}\left(  \theta\mathbf{E}\right)  +\frac{\alpha}{\left(  2\pi\right)
^{2}}\frac{\partial}{\partial t}\left(  \theta\mathbf{B}\right)
=\frac{c\alpha}{\left(  2\pi\right)  ^{2}}\left(  \mathbf{\nabla}%
\theta\mathbf{\times E+}\theta\mathbf{\nabla\times E}\right)  +\frac{\alpha
}{\left(  2\pi\right)  ^{2}}\left(  \theta\frac{\partial\mathbf{B}}{\partial
t}+\mathbf{B}\frac{\partial\theta}{\partial t}\right)  $.

In particular, note the continuity equation for the $\theta$-terms, namely
$\mathbf{\nabla\cdot J}_{\theta}+\frac{\partial\rho_{\theta}}{\partial t}=0$
\ (this basically reflecting the conserved \textquotedblleft Witten electric
current\textquotedblright\ $-$ see mention of the Witten effect further below).

As is well-known, the above originate from an extra term in the
electromagnetic Lagrangian density, that is of the form $\pounds _{axion}%
=\theta\left(  \frac{e^{2}}{2\pi hc}\right)  \mathbf{E\cdot B}=\theta
\frac{\alpha}{(2\pi)^{2}}\mathbf{E\cdot B}$, which is periodic with respect to
$\theta$ with period $2\pi$ (and if $\theta$ is static, it only takes values
$0$ or $\pi$ for t-reversal-symmetric systems $-$ $0$ being the value for
conventional, and $\pi$ for topologically nontrivial systems). The proof of
this periodicity that has been recently given by Vazifeh \&
Franz\cite{VazifehFranz} in the absence of magnetic monopoles can actually be
given directly by our natural criterion, by actually justifying better the
\textit{separate }quantization of certain fluxes that does not seem to be
justified in ref.\cite{VazifehFranz}. For the example of \cite{VazifehFranz}
with $\mathbf{B=}B_{z}\hat{e}_{z}$ and $\mathbf{E=}E_{z}\hat{e}_{z}$, we see
that our nonlocal fluxes appear naturally after integration (over spatial and
time variables) of $\pounds _{axion}$ in order to obtain the axionic action
$S_{axion}$, namely $S_{axion}/\hbar=\frac{\theta}{\Phi_{0}^{2}}\int
B_{z}dxdy\int E_{z}cdtdz.$ And although the separate quantization of the
fluxes in eq.(16) of ref.\cite{VazifehFranz} does not seem to result from any
basic principle, it \textit{is} justifiable by our quantization of nonlocal
terms (the one that appears above with the $B_{z}$, a usual magnetic flux, and
the one with $E_{z}$, a spacetime electric flux). So the separate quantization
gives $n_{1}\Phi_{0}$ for the magnetic flux and $n_{2}\Phi_{0}$ for the
electric flux ($n_{1},n_{2}$ integers), so that finally $S_{axion}%
/\hbar=N\theta$ ($N=n_{1}n_{2}$) as we were seeking to prove. In a sense, the
above separate quantization of magnetic and electric fluxes proves that the
axionic action (which is a 2nd Chern number for this Abelian gauge theory)
turns out to be a product of two 1st Chern numbers, whose quantization comes
out directly by imposing our criterion of proper behavior (i.e. enforcing the
elimination of gauge ambiguity in the planes $(xy)$ and $(tz)$), and without
further topological considerations.

\bigskip

Returning to the above generalized Maxwell's equations, note that
$\int\mathbf{\nabla\cdot J}_{\theta}dt=-\rho_{\theta}=$ $\frac{\alpha}{\left(
2\pi\right)  ^{2}}\mathbf{\nabla\cdot}\left(  \theta\mathbf{B}\right)  $, so
that its volume integral in a spatial region will give a flux of
$\theta\mathbf{B}$ through the surface boundary. And because $\theta$ just
suffers a jump by $\pi$ at the surface (if this is the interface between an
axionic medium and a conventional one, i.e. the vacuum) there remains just the
magnetic flux through the surface; hence our criterion of its flux
quantization leads naturally to the quantization of $\mathbf{J}_{\theta}$.
However, as has been noted in the past as an observation, $\mathbf{J}_{\theta
}$ can describe the Hall current, $\mathbf{J}_{\theta}=\mathbf{J}_{Hall}%
$\textbf{ (}see below for a new and clear proof), hence the above conclusion
on $\mathbf{J}_{\theta}$ leads to an immediate understanding of the
quantization of the Hall response. But what type of quantization? (It will
turn out in the following that there are two types, integral and
\textquotedblleft half-integral\textquotedblright).

\bigskip Let us first consider a conventional 2D Quantum Hall (QH) sheet (i.e.
with $\sigma_{_{H}}=-\nu e^{2}/h$, with $\nu$ an integer (the filling factor
in a Landau level picture $-$ or, more generally, the 1st Chern number in the
Brillouin zone of a periodic system)) and let us fold it along the
$x$-direction into a cylinder. Applying an electric field $E$ on the surface,
with $E$ parallel to the cylinder axis, we obtain a transverse Hall current
$I$ (hence in the azimuthal direction) of density $J=\sigma_{_{H}}E$, where
$J=I/d$ (with $d$ the height of the cylinder). We have therefore a total
magnetic moment $M_{tot}$ induced that is parallel to the $z$-axis and has a
value $M_{tot}=-\frac{IS}{c}$ (this comes out if we view $I$ as related to the
magnetization current through $\mathbf{J}=\mathbf{J}_{magn}=c\mathbf{\nabla
}\times\mathbf{M}$) $\Rightarrow$ $\mathbf{\tilde{n}}\times(\mathbf{M}%
_{2}-\mathbf{M}_{1})=\frac{1}{c}\mathbf{K}_{magn}$, with $\mathbf{K}_{magn}$
the surface current per transverse length, and with the cross section $S=\pi
R^{2}$, so that $M_{tot}=-\sigma_{_{H}}\frac{S}{c}dE=-\nu\frac{e^{2}}{h}%
\frac{V}{c}E$ (where we wrote $Sd=V$ the volume of the cylinder). We have
therefore an induced (by the electric field) magnetization $M=\frac{M_{tot}%
}{V}=-\nu\frac{e^{2}}{hc}E$. Now, it is well-known (and it results from
variation of the above $S_{axion}$) that axionic physics leads to
magnetoelectric effects such that an electric field $\mathbf{E}$ induces a
parallel magnetization $\mathbf{M}=\frac{\alpha}{4\pi}\frac{\theta}{\pi
}\mathbf{E}$, whereas a magnetic field $\mathbf{B}$ induces a parallel
polarization $\mathbf{P}=\frac{\alpha}{4\pi}\frac{\theta}{\pi}\mathbf{B}$. The
above picture of the IQHE already leads to (recall that $\alpha=e^{2}/\hbar
c=2\pi\frac{e^{2}}{hc}$) a magnetization $M=-\nu\frac{e^{2}}{hc}E=-\nu
\frac{\alpha}{2\pi}E$ which corresponds to the above general result with
$\theta=2\pi\nu$. (This probably demonstrates in a sense the conventional
character of the IQHE). We can then actually show that this correspondence
($\theta=2\pi\nu$) is also valid for the polarization induced by a magnetic
field. Indeed, by applying a $\mathbf{B}$ in all space parallel to the
cylinder axis we obtain as a response an electric charge density, say $n_{A}e$
($n_{A}$ being the areal number density of charge carriers) induced on the
ceiling of the cylinder $-$ and one of the same magnitude induced on the
floor, but with opposite sign (hence we \textit{do} indeed have a dipole
electric moment and therefore a polarization induced, parallel to the axis)
$-$ a result that comes from the well-known Streda formula\cite{Streda}
$\sigma_{_{H}}=-\frac{\partial\left(  en_{A}c\right)  }{\partial B}$ (see
below), and the $B-n_{A}$ \textquotedblleft locking formula\textquotedblright%
\ of the IQHE, namely $B=\frac{n_{A}\Phi_{0}}{\nu}$ (that connects the
constant (and very robust) value of $B$ at the plateau labeled by the integer
$\nu$ in an IQHE experiment), or in an even more elementary manner by the
polarization charge density $\rho_{pol}=-\mathbf{\nabla\cdot P}$ $\Rightarrow$
$(\mathbf{P}_{2}-\mathbf{P}_{1})\cdot$ $\mathbf{\tilde{n}}$ $=-\sigma_{pol}$
the surface polarization charge density induced on each considered surface.
Let us check whether this polarization response satisfies the above general
axionic expression, again with $\theta=2\pi\nu\mathbf{:}$ indeed, from Streda
formula $\sigma_{_{H}}=-\frac{\partial\left(  en_{A}c\right)  }{\partial B}$
together with $n_{A}=\frac{\nu B}{\Phi_{0}}$ we get $\sigma_{_{H}}%
=-\frac{\partial}{\partial B}\left(  e\frac{\nu B}{\frac{hc}{e}}c\right)
=-\nu\frac{e^{2}}{h}$ the correct Hall conductance, and therefore the total
electric moment induced in the $z$-direction is $P_{tot}=-en_{A}Sd=-e\frac{\nu
B}{\frac{hc}{e}}V$; hence finally the polarization is $P=\frac{P_{tot}}%
{V}=-\nu\frac{e^{2}}{hc}B=-\nu\frac{\alpha}{2\pi}B$, and we see that this
indeed also satisfies the above general axionic response relation, with
$\theta=2\pi\nu$.

Along similar lines one can find in the literature\cite{refaxion} the manner
in which one can obtain the anomalous (half-integral) Quantum Hall Effect at
the surface of a strong topological insulator, by using an argument such as
the above, with $\theta=\pi$ for the topological insulator and $\theta=0$ for
vacuum (hence the above spatial derivatives of $\theta$ give delta function
contributions (from $\rho_{\theta}=-\frac{e^{2}}{2\pi h}\mathbf{\nabla}%
\theta(\mathbf{r})\mathbf{n\cdot B}$ and $\mathbf{J}=c\frac{e^{2}}{2\pi
h}\mathbf{\nabla}\theta(\mathbf{r})\times\mathbf{E}$ at the interface, as we
move from one medium to the other)). Indeed, for a flat 2D system one obtains
$\rho_{\theta}=\frac{\alpha}{4\pi^{2}}\delta(z)B_{z}$ and $\mathbf{J}_{\theta
}=-\frac{\alpha}{4\pi^{2}}\delta(z)\hat{e}_{z}\times\mathbf{E}$ (for each
surface of a topological insulator film) and it finally turns
out\cite{refaxion2} that $\sigma_{_{H}}=\left(  \theta_{1}-\theta_{2}+2\pi
n\right)  \frac{e^{2}}{2\pi h}$, leading (i.e. for $\theta_{1}=\pi$ and
$\theta_{2}=0$) to the \textquotedblleft half-integral\textquotedblright%
\ quantization for topological insulators, namely $\sigma_{_{H}}=\frac{e^{2}%
}{2h}$ modulo $\frac{e^{2}}{h},$ or simply $\sigma_{_{H}}=$ $\pm\frac{e^{2}%
}{2h}$. And this, applied to the surface of a topological insulator cylinder
(in a way similar to our above application to an IQHE system) also gives rise
to a quantized magnetoelectric response $\mathbf{M}=\frac{\alpha}{4\pi
}\mathbf{E}$ and $\mathbf{P}=\frac{\alpha}{4\pi}\mathbf{B}$, compatible with
the above discussion and quite generally expected for topologically nontrivial
materials (corresponding to $\theta=\pi$). Note that deep down, the origin of
the above is essentially the Witten effect\cite{Witten}, which is usually
presented as follows\cite{Franz}: from $\mathbf{\nabla\cdot E=}4\pi\left(
\rho+\rho_{\theta}\right)  $ and considering now $\theta$ to be a function of
time only $\theta(t)$, and using the possible presence of magnetic monopoles
through $\mathbf{\nabla\cdot B=}4\pi\rho_{m}$, it turns out that
$\int\mathbf{\nabla\cdot E}d^{3}r=4\pi\int d^{3}r\left(  \rho(\mathbf{r}%
)-\frac{\theta}{\pi}\alpha\rho_{m}(\mathbf{r})\right)  $; hence if the total
charge is $Q=\int d^{3}r\rho(\mathbf{r})=-Ne$, and $\theta=\pi$, and because
$4\pi\int\rho_{m}d^{3}r=n\Phi_{0}$ (the Dirac quantization, as derived
earlier), we get $\int\alpha\rho_{m}d^{3}r=ne/2$, or that effectively we have
a charge $Q_{eff}=-e\left(  N-\frac{n}{2}\right)  $, namely an effective
charge that is generally half-integral\cite{Witten}. Summarizing, we see that
imposition of our criterion of proper behavior (that enforces elimination of
our gauge ambiguity) leads to quantization of $\mathbf{J}_{\theta}$, which in
turn leads, for conventional IQHE systems to $\sigma_{_{H}}=$ an integral
multiple of $\frac{e^{2}}{h}$, and for topological insulator surfaces that are
in contact with a topologically trivial medium (i.e. the vacuum) to
$\sigma_{_{H}}=$ an odd integral multiple of $\frac{e^{2}}{2h}$; it also leads
to their quantized magnetoelectric responses, as these were discussed above.

One can actually generalize the above magnetoelectric effects to more general
topologically nontrivial quantum devices, combined with the field-effect from
a distance (the central result of the main article), but this is reserved for
a future discussion.

\subsection{Predictions on devices, Graphene and Topological Insulators}

Descriptions of possible types of measurements (related to the gauge
nonlocality effect presented in this work) in conventional systems have been
briefly given in the main text and in Appendix E (mostly on induction of IQHE
from outside the system). It should be stressed, as a generic feature (and
prediction) that, even if our \textquotedblleft white area\textquotedblright%
\ is empty (i.e. single electron in empty space), we would at least expect
(persistent) currents along the edge (interface between white and dark areas)
$-$ this being valid for both parabolic and Dirac electronic
spectrum\cite{Sticlet}. This was also noted in the main text (with the
expectation that $J$ will be proportional to $\partial\epsilon/\partial y_{0}$).

Here is what one would expect on general grounds, if our white system is
graphene or a topological insulator$\mathbf{:}$ \textit{Graphene}\textbf{:}
proximity arrangement with a $\mathbf{B}$, would offer a controllable way
(through changes of the outside $\mathbf{B}$ or of $y_{0}$) to \textit{lift}
the \textit{orbital degeneracy} that originates from the two valleys (even
without inter-valley scattering), with consequences on persistent currents (in
$x$-direction) and in conductance (i.e. some shifting of peaks), analogous to
the ones of ref.\cite{graphene1}. In addition, giant magnetoresistance at room
temperature is possible, due to the hidden AB interference\cite{graphene2}.
\textit{Topological insulators}\textbf{:} By way of an example, in the
proximity of an HgTe quantum well one would expect to measure helical edge
states, bound states and persistent currents (with Rashba spin-orbit
coupling), that would generally be affected in a manner similar to the one
described in ref.\cite{TI}. On all this, we plan to return with details in a
future note.

\subsection{Formal analogies with other areas}

Finally, the purpose here is to examine the wider physical implications,
and/or relationships with other physics areas that we see \textit{formal}
connections with.

1. For time-dependent fields, the \textit{time-derivative} of the
phase-nonlocalities noted in the present work (and expressed through the
nonlocal terms of Appendix B) seems to be directly related to recent
considerations of Berry and Shukla\cite{BerryShukla} on \textquotedblleft curl
forces\textquotedblright\ that are spatially confined in classical systems
(while the point of observation is \textit{outside}, in curl-free regions),
giving simultaneously their quantum generalization (to be addressed in
separate work). A \textquotedblleft\emph{curious evocation of the AB
effect}\textquotedblright\ is a statement mentioned twice in
\cite{BerryShukla}, and it will be shown that this is actually related to our
nonlocal terms, and to the nonlocal gauge influences that generally show up in
the spirit of the present work (and more concretely, to the hidden AB effect
in some 3D companion system).

2. For a many-body system of non-interacting electrons there will be
interesting transitions (upon variation of the width $d$ of the
\textquotedblleft white\textquotedblright\ area), such as the ones recently
worked out in detail in 3D systems\cite{EPJB} (note in particular that there
is a gap opening due to the nonzero values of $k_{y}$ and this will have
consequences). The above is for our simple \textquotedblleft
empty\textquotedblright\ problem, but can also be worked out in a
\textquotedblleft white\textquotedblright\ graphene sample (a Dirac material,
outside a nearby field, now with electron and hole bands taken into account).

3. An obvious generalization of these proximity influences to a many-body
system with electron-electron interactions (but with no magnetic field inside
our system) leads to another novel possibility, of potential relevance to the
physics of composite fermions \textit{without} the presence of a magnetic
field (where the extra vector potentials induced by proximity can lead to
corresponding Chern-Simons physics, in a manner similar to the one studied
recently in ref.\cite{Gvozdikov}).

\bigskip

4. Possible applicability to other systems (i.e. that involve rotations rather
than magnetic field (due to the well--known formal similarity between the two
physical situations) and that give rise to analogies with the AB effect, with
i.e. water waves, pioneered by Berry\cite{BerryAB}): such systems have been
recently noted to exhibit quantization of orbits as well as nonlocality
(generated by path memory) $-$ see the very recent preprint \cite{Fort}, and
the gauge nonlocality advanced in the present work might be directly applicable.

\bigskip

5. Spin-physics, through a combination of remote field influences with proper
Lorentz boosts: it is well-known (see i.e. ref.\cite{Xu}) that in setups such
as the ones discussed in the main text (i.e. the initial example of an
orthogonal strip), if one boosts to a moving frame with velocity
$\mathbf{v}=c\frac{\mathbf{E}\times\mathbf{B}}{E^{2}}$, then, as a result, the
moving observer perceives a magnetic moment $\mathbf{%
%TCIMACRO{\U{3bc} }%
%BeginExpansion
\mu
%EndExpansion
}=\frac{1}{2}q\mathbf{r}\times\mathbf{v}$ for the particle. Using then
$\mathbf{B}=\frac{1}{c}\mathbf{v}\times\mathbf{E}$, it turns out that the
moving observer experiences the vector potential as $\frac{q\mathbf{A}}%
{c}=\frac{1}{c^{2}}\mathbf{%
%TCIMACRO{\U{3bc} }%
%BeginExpansion
\mu
%EndExpansion
\times E}$, namely a vector potential of the Aharonov-Casher type\cite{AC}.
All the earlier quantization conclusions can be therefore transported to
quantization and IQHE-type of phenomena that concern the particle's magnetic
moment and spin. We shall return to this in a future article, but for here it
suffices to note that nontrivial spin-physics can be studied this way,
starting from purely orbital considerations.

\bigskip

6. One cannot stop wondering whether the new nonlocal terms (basically
responsible for the above effects (see Appendix B), viewed as generalized
solutions of $\mathbf{\nabla}\Lambda=\mathbf{A}$ on the simple-connected
plane, but \textit{in problems where }$\Lambda$\textit{ is not defined
everywhere as a single-valued function}) can have an impact on other areas of
physics where we have such partial differential equations on a plane. One
example of applicability that we already saw is the Berry \& Shukla
problem\cite{BerryShukla} mentioned earlier in this Appendix. Another
immediate candidate is the entire area of thermodynamics, filled with 1st
order partial differential equations of \ this form, and it seems that,
indeed, the nonlocal terms that appear in their solutions might have
connection to issues of irreversibity and vorticity\cite{Kiehn} that, for now,
go much further than the scope of this paper.

\bigskip

\bigskip\ 7. Finally, and again going much beyond the scope of the present
article, one cannot help noticing that there seems to be a general connection
of the above with certain high-energy physics phenomena. Although the physics
is very different (the dynamical variables involved also being different),
there seems to be a \textit{formal} relationship that might be useful (at
least through analogies). Such a formal analogy is the recent work in Maxwell
electrodynamics on a compact manifold, that finds a topological contribution
to the Casimir force\cite{Zhitnitsky} that seems to be the formal analog of
the persistent currents asserted here (upon variation of $y_{0}$) due to the
proximity effect of the present work. More generally, note earlier works on
$\theta$-vacuum (see i.e. ref.\cite{Asorey}) with $\theta$-vacuum sectors
being formally analogous to our $y_{0}$--sectors (which $-$ if they were
dynamic (i.e. if the 3D $\mathbf{B}$-construction or the cylinder had their
own dynamics (i.e. a vibrational one)) $-$ they would be analogous to axions).
In fact, in ref.\cite{Zhitnitsky} on topologically inequivalent (winding)
states, where, in addition, a direct connection of $\theta$-vacua with
magnetic fields is made, one can see a similar formal analogy with our simpler
system; in a sense, our work points out to another \textquotedblleft amazing
example\textquotedblright\ (in the language of the authors of
ref.\cite{Zhitnitsky}) $-$ now in low-energy physics $-$ where the external
$B$ is now outside the system (playing the role of the $\theta$-parameter)).
Similarly, one can see some possible formal analogy of our $y_{0}$-ambiguity
to the so-called Gribov problem (or Gribov ambiguity\cite{Gribov}). This will
be valid, when the Gribov copies are gauge-equivalent configurations that
satisfy the Landau gauge condition\cite{Lavrov}. For such a claimed connection
see in particular refs \cite{Semenoff} and \cite{Hetrick} where the existence
of the Gribov phenomenon is related to the existence of inequivalent
quantizations (which in our simpler problem means different $y_{0}$-sectors),
and therefore Gribov copies are labeled through procedures that are formally
similar to ours; and this is done for \textit{abelian} gauge fields at
\textit{zero} temperature (as noted in ref.\cite{Pramana}) - although
originally Gribov copies were discovered for only non-abelian gauge fields at
a general finite temperature. In particular there is an issue of the Gribov
problem showing up when gauge fields (like our potentials) do not vanish at
infinity $-$ and it seems that a similar issue that has gone largely unnoticed
exists here as well: i.e. in the majority of literature on AB effects (see as
an example the very recent work of Stewart\cite{Stewart}) there is a proof in
Coulomb gauge that there is no further gauge ambiguity, the reason being the
vanishing at infinity of vector potential $A$ in a standard AB configuration
with cylindrical symmetry (in which case, $A$ goes as $1/\varrho$ outside the
enclosed flux and indeed vanishes at infinity); this however is not true if
the inaccessible region is i.e. a rectangle (and it is not true for our
striped geometry either), because in such orthogonal geometries (where our
gauge can be used) the corresponding $A$ outside the field is a nonzero
constant that can go up to infinity.\ Hence a basic assumption, usually made
implicitly, does not hold in our case (and mathematically speaking it leads to
the gauge ambiguity emphasized in the present work), and something formally
similar seems to hold in high energy physics\cite{Ellis} for systems that
exhibit the Gribov ambiguity. Needless to say, all these issues are only
briefly mentioned here without justification and require closer scrutiny;
however, we feel it is useful to point them out, in case that the gauge
nonlocality and the associated proximity effect found in the present work,
might be the low-energy analog of previously known (but mostly esoteric)
technical matters in high-energy physics, contributing therefore to their
demystification (and, possibly, vice versa: bringing out some esoteric and
sophisticated behavioral patterns, that may be hidden in an \textquotedblleft
ordinary\textquotedblright\ solid state system in the laboratory, and with the
actual possibility of experimental and practical applications). Such analogies
would be useful for possibly making further progress in deeper gauge-related
issues in both high- and low-energy physics, and this is why they deserve to
be investigated further.

\end{document}